\documentclass[letterpaper, 10 pt,conference]{ieeeconf}  

\IEEEoverridecommandlockouts                              

\usepackage{xcolor}
\usepackage{graphicx}
\usepackage{subcaption}
\usepackage{amsmath}

\graphicspath{{./}{./images/}}

\usepackage{color}
\usepackage{hyperref}
\definecolor{linkcolor}{RGB}{83, 83, 192}
\hypersetup{colorlinks, linkcolor=linkcolor}

\usepackage{algpseudocode, algorithm}

\title{\LARGE \bf
Extraction of Nystagmus Patterns from Eye-Tracker Data with Convolutional Sparse Coding.
}

\author{Clément Lalanne$^{a}$, Maxence Rateaux$^{b}$, Laurent Oudre$^{c}$, Matthieu P. Robert$^{b}$, Thomas Moreau$^{d}$
\thanks{$^{a}$ Clément Lalanne is with Centre Borelli, CNRS, ENS Paris Saclay, Université de Paris, SSA.}%
\thanks{$^{b}$ Maxence Rateaux and Matthieu P. Robert are with Hôpital Necker-Enfants Malades, AP-HP, and Centre Borelli, CNRS, ENS Paris Saclay, Université de Paris, SSA.}%
\thanks{$^{c}$ Laurent Oudre is with Université Sorbonne Paris Nord, L2TI, UR 4443, F-93430, Villetaneuse, France}%
\thanks{$^{d}$ Thomas Moreau is with University Paris-Saclay, Inria, CEA {\tt\small thomas.moreau@inria.fr}.}%
}

\begin{document}

\maketitle
\thispagestyle{empty}
\pagestyle{empty}

\begin{abstract}
The analysis of the Nystagmus waveforms from eye-tracking records is crucial for the clinicial interpretation of this pathological movement. A major issue to automatize this analysis is the presence of natural eye movements and eye blink artefacts that are mixed with the signal of interest. We propose a method based on Convolutional Dictionary Learning that is able to automaticcaly highlight the Nystagmus waveforms, separating the natural motion from the pathological movements. We show on simulated signals that our method can indeed improve the pattern recovery rate and provide clinical examples to illustrate how this algorithm performs.
\end{abstract}

\section{INTRODUCTION}

Nystagmus is a variety of involuntary eye movement, acquired in infancy or later in life, that may be associated with reduced or limited vision. It is characterized by a periodic movement of one or both eyes. The many types of Nystagmus can be roughly classified in two large classes of motions. The first ones, referred to as \textit{pendular} Nystagmus, manifest as slow periodic oscillations. The second ones, referred to as \textit{jerk} Nystagmus, are characterized by the alternation of slow phases and quick phases corresponding to eye saccades. In total, Hertle et al. \cite{hertle_differential_2013} reported 49 differents types of Nystagmus and 17 types of saccadic intrusions and oscillations; classified in 8 main categories according to the Classification of Eye Movement Abnormalities and Strabismus (CEMAS)\cite{hertle_next_2002}. Several etiological conditions may explain the presence of Nystagmus, such as the presence of an infantile strabismus, a developemental disturbance of the afferent visual pathways or of the oculomotor sytem. However, there is currently no medical consensus on the diseases that are linked to a specific type of Nystagmus. It is therefore an active research question: the automatic classification and characterization of this phenomenon could improve the diagnosis, prevention and follow-up of patients suffering from this condition.

\begin{figure}[t]
    \centering
    \includegraphics[width=0.8\columnwidth]{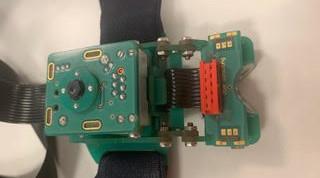}
    \caption{EyeFant Tracker}
    \label{fig:tracker}
\end{figure}

Traditionally, the detection and monitoring of Nystagmus on young infants was performed by specialized ophthalmologists who study the eye movements by visual inspection. Recently, some dedicated sensors were developed to track the position of both eyes with respect to time (see \autoref{fig:tracker}). The EyeFant Tracker, designed by Ober Consulting\footnote{http://www.ober-consulting.com/}, has been specifically conceived for eye movement recording in infants. The sensors -- located in the shadow of the nose -- localize the minimum of infra-red cornea reflexion of the eye and deduce the position of the pupil from it. The signal are sampled at 1kHz with a spatial resolution of 0.1$^{\circ}$ and a linear range of $\pm$35$^{\circ}$ horizontally and $\pm$20$^{\circ}$ vertically. They output left- and right-eye trajectories in the 2D plane which can be visualized with 4 graph: vertical and horizontal components for the each eye.

\begin{figure}[t]
    \centering
    \includegraphics[width=0.8\columnwidth]{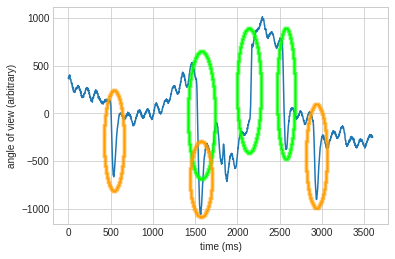}
    \caption{Example of eye tracker recording of a patient with a pendular Nystagmus (left-eye, horizontal). The signal includes saccades (\emph{green}) as well as eye blinks artefacts (\emph{orange}) hindering the automatic extraction of Nystagmus waveforms.}
    \label{fig:traj}
\end{figure}
\autoref{fig:traj} displays a typical eye-trajectory in one axis. This record comes from one eye of a patient with a pendular Nystagmus. It shows the horizontal projection of the angle that makes the left eye with a normal vector to the face plan. Three simultaneous phenomena are actually recorded by the eye tracker: the low-frequency component is due to the natural eye movement, the abrupt changes correspond to saccades and eye blinks artefacts (in green and orange) and the small oscillations are due to the Nystagmus (here periodic).  The aim of this article is to automatically extract from such signals the components linked to the Nystagmus as patterns associated with activations. This task is especially tricky since it requires to be robust to the other phenomena, i.e. saccades, artefacts and eye movements. To that aim, we propose to use a machine learning tool known as convolutional sparse coding \cite{Grosse2007} and to adapt it to deal with signals corrupted by a trend (eye movements) and ruptures (saccades and artefacts).

\section{BACKGROUND}

Convolutional sparse coding aims at representing an input signal as the convolution between  patterns (that capture the phenomenon present in the signals) and activations (that specify when these patterns appear in the signal).

More specifically, given a input signal $\mathbf{x}$ of length $T$, the aim is to learn a library of $K$ patterns $\mathbf{D} = (\mathbf{d_k})_{1\leq k \leq K}$ of length $W$ (with $W<<T$) and their associated activations $\mathbf{Z} = (\mathbf{z_k})_{1\leq k \leq K}$ of length $L = T-W+1$. The general model writes as
\begin{equation}
    \mathbf{x} = \sum_{k=1}^{K} \mathbf{d_k} * \mathbf{z_k} + \mathbf{e}
             = \mathbf{D} * \mathbf{Z} + \mathbf e
    \enspace,
    \label{eq:model}
\end{equation}
where $*$ is the discrete convolution product and $\mathbf{e}$ a noise term \cite{Grosse2007}.

With matrix notations, the Convolutional Sparse Coding  (CSC) problem rewrites as
\begin{equation}
    \underset{\substack{\mathbf{D},\mathbf{Z}\\\forall k, ||D_k||_2^2 \leq 1}}{\text{min}}\  \frac{1}{2}||\mathbf{D} * \mathbf{Z} - \mathbf{x}||_2^2 + \lambda ||\mathbf{Z}||_1,
    \label{eq:prob}
\end{equation}
where $\lambda$ allows to control the level of sparsity put on the activations.

This problem is usually solved through alternate minimization, where matrix $\mathbf{Z}$ is updated while $\mathbf{D}$ is fixed (\textsc{sparseCoding}), and reciprocally (\textsc{dictUpdate}). Popular solvers for this problem include AlphaCSC \cite{la2018multivariate} and Sporco \cite{Wohlberg2016}. It has been used in several contexts related to healthcare such as blood cell detection in holographic imaging \cite{Yellin2017} or to measure respiration and heartbeat rates in EEG signals \cite{Wang2019a}.

\section{PROPOSED METHOD}

In our context, the input signal $\mathbf{x}$ corresponds to an eye trajectory in horizontal or vertical direction, and we aim to learn from this signal a series of recurring patterns stored in $\mathbf{D}$ that would correspond to Nystagmus waveforms and their temporal activations stored in $\mathbf{Z}$. However, as seen on \autoref{fig:traj}, two types of eye movements limit the use of CSC in this context. Indeed, the input signal $\mathbf{x}$ does not only include the Nystagmus phenomenon, but also the voluntary eye movements, that manifests as a smooth motion and a few eye saccades. Hence, the input signal presents abrupt ruptures that are not taken into account in the model \eqref{eq:model}. The aim of the proposed model is to modify the standard CSC problem \eqref{eq:model} so as to deal with smooth trends and abrupt ruptures.

\subsection{Convolutional Sparse Coding with Piecewise Constant Trend}

The Convolutional Sparse Coding with Piecewise Constant Trend (CSC-PCT) model introduced in this article consists in adding an additional term into the standard CSC model:
\begin{equation}
    \mathbf{x} = \mathbf{D} * \mathbf{Z} + \mathbf{y} + \mathbf{e},
    \label{eq:model2}
\end{equation}
where $\mathbf{y}$ corresponds to a piecewise constant trend term.
The associated optimization problem writes as
\begin{equation}
    \underset{\substack{\mathbf{D},\mathbf{Z},\mathbf{y}\\\forall k, ||D_k||_2^2 \leq 1}}{\text{min}}\  \frac{1}{2}||\mathbf{D} * \mathbf{Z} + \mathbf{y} - \mathbf{x}||_2^2 + \lambda ||\mathbf{Z}||_1 + \lambda_{TV} || \nabla \mathbf{y} ||_1,
    \label{eq:prob2}
\end{equation}
where $|| \nabla \mathbf{y} ||_1$ is a  Total Variation (TVL1) constraint defined as
\begin{equation}
   || \nabla \mathbf{y} ||_1 = \sum_{t} |y_{t+1} - y_t|.
\end{equation}
The additional constraint imposes a piecewise constant structure to $\mathbf{y}$, which allows the modelling of the abrupt ruptures as well as the changes in baseline values due to low-frequency phenomenons.

\subsection{Resolution}

In order to solve \eqref{eq:prob2} with respect to $\mathbf{D}, \mathbf{Z}, \mathbf{y}$, the problem is divided into 3 sub-problems that each updates one term while keeping the remaining terms fixed. These steps are summarized in \autoref{alg:detrend_csc}.

Note that the $\mathbf{D}$- and $\mathbf{Z}$-updates are actually the same than in classical CSC, by replacing $\mathbf{x}$ in \eqref{eq:prob} by $\mathbf{x}-\mathbf{y}^{(q)}$, where $\mathbf{y}^{(q)}$ is the trend estimation at the $q^{th}$ iteration. Therefore, classical solvers can be used. We relied on AlphaCSC solver \cite{la2018multivariate} which uses a coordinate descent approach for \textsc{sparseCoding} and FISTA for \textsc{dictUpdate}.

For the $\mathbf{y}$-update, the aim is to compute a piecewise estimate $\mathbf{y}^{(q)}$ of the current residual $\mathbf{x} - \mathbf{D}^{(q)} * \mathbf{Z}^{(q)}$. This corresponds to applying the proximal operator of TVL1 to the residual, for which several off-the-shelf solvers are available. In this article, we used \textsc{ProxTV} \cite{JMLR:v19:13-538} which computes the update through a dual framework.

\begin{figure}
\vspace{-.5em}
\end{figure}
\begin{algorithm}[t]
    \begin{algorithmic}[1]
        \Require $x, \pmb D^{(0)}$, parameter $\lambda >  0, \lambda_{TV} > 0$,
                and $\epsilon > 0$.
        \State $y^{(0)} = \textsc{proxTV}(x, \lambda_{TV})$
        \Repeat
            \State $Z^{(q+1)} = \textsc{sparseCoding}(X - y^{(q)}, D^{(q)}, \lambda)$
            \State $y^{(q+1)} = \textsc{proxTV}(x - Z^{(q+1)} * D^{(q)}, \lambda_{TV})$
            \State $D^{(q+1)} = \textsc{dictUpdate}(x - y^{(q)}, Z^{(q)}, \lambda)$
        \Until{$\|Z^{(q+1)} - Z^{(q)}\|_\infty < \epsilon$}
    \end{algorithmic}
    \caption{Locally Greedy Coordinate Descent}
    \label{alg:detrend_csc}
\end{algorithm}

\subsection{Scalability to other setups}

The problem \eqref{eq:prob2} uses a classical method in optimization called penalization. Namely, it includes the constraints (smooth trend with saccades) in the function to optimize by measuring "how far" the result is from the constraints set. Here, this is done with the term $|| \nabla \mathbf{y} ||_1$.
As a consequence, our algorithm can be extended to any setup where the motion to get rid of can be expressed by a proximable penalization term.

\section{DATA}

\subsection{Synthetic dataset}
\label{sub:simulated_data}

To evaluate our framework, we used the generative model \autoref{eq:model2} to simulate look-a-like signals for our data. The voluntary gaze movements are created as the sum of a low frequency pattern and saccades that are sigmoidal ruptures and hence allow abrupt yet smooth mode shifting. The localisation and amplitude of the saccades are chosen randomly with a Poisson law of parameter 1s and a mean amplitude of 20$^\circ$. The Nystagmus waveforms are generated according to the rough existing Nystagmus classification. The pendular Nystagmuses are modelled by a sinusoidal signal and the jerks are modelled by quadratic slow phases and linear quick phases. The frequency and amplitude of the Nystagmus are also chosen randomly with mean 5Hz and 3$^\circ$. The resulting signal $x$ is obtained by summing these two components. The generated signals have a sampling rate of 1000Hz and length of 10s and a white noise of $0.5^\circ$ is added to match visually the signals obtained with our recording device.

\subsection{Real dataset}
\label{realdataset}

Our clinical evaluation was perfomed on two different eye movements recordings from two different patients. The first one was realized  on an 6-years-old girl with a Down syndrome. The patient exhibited a \textit{Spasmus Nutans type-Nystagmus} (SNTN). There was no ocular deviation and the refraction was: -6.25 (+1.75 80$^{\circ}$) for the right eye (RE) and -5.00 (+1.25 110$^{\circ}$) for the left eye (LE). The Nystagmus was monocular (only in the RE) and purely pendular. There was no head nodding and no head turn.
The second eye movement recording was performed on a 36-years-old women with an oculocutaneous albinism (AOC) and an infantile Nystagmus syndrome (INS). The best corrected visual acuity was +0.7 logMAR (RE and LE) and the refraction was: -7.25 (+4.00 80$^{\circ}$)/ -8.75 (+3.25 100$^{\circ}$). Ophthalmological examination revealed an iris transillumination, a peripheral hypopigmentation of the fundus and a foveal hypoplasia.
 Consents were collected in both cases, according to the French regulation about personal data.

\section{RESULTS}

All experiments are performed in Python and the code is available online\footnote{\url{https://github.com/tommoral/detrending_cs_oculo}}

\subsection{Pattern recovery on synthetic signals}

To evaluate the benefit of joint detrending with CSC on pattern discovery, we used 160 signals generated according to the procedure described in \autoref{sub:simulated_data} with random parameters.
For each signal $x_i$, we learn $K=1$ atom $d_i$ using either our method (\textsc{Joint}) or the vanilla CSC method with either no pre-processing (\textsc{None}) or an initial detrending step (\textsc{Init}). We then evaluate the recovery of the pattern $\mathbf{d}_i^*$ used to generate the Nystagmus part. Due to the periodicity of the Nystagmus signal, the learned atoms might have a different phase compared to the original pattern and capture multiple periods. To make sure we correctly capture the pattern with the learned atoms, we used the following metric
\[
    \rho(\mathbf{d}^*, \mathbf{d}) = \max_{\substack{t=1..L\\l=1..W - L}}
        \frac{<\mathbf{\widetilde d}^*[t:t+L], \mathbf{d}[l:l+L]>}{\|\mathbf{d}^*\|_2\|\mathbf{d}\|_2}
\]
where $\mathbf{\widetilde d}$ is the periodized version of $\mathbf{d}$ and $\mathbf{d}[t_1:t_2]$ denotes the coefficient from $t_1$ to $t_2$ in $\mathbf{d}$. This metric extends the Pearson coefficient to be invariant to the phase and the repetition of $\mathbf{d}^*$ in the learned atom $\mathbf{d}$.

\begin{figure}[tp!]
    \centering
    \includegraphics[width=\columnwidth]{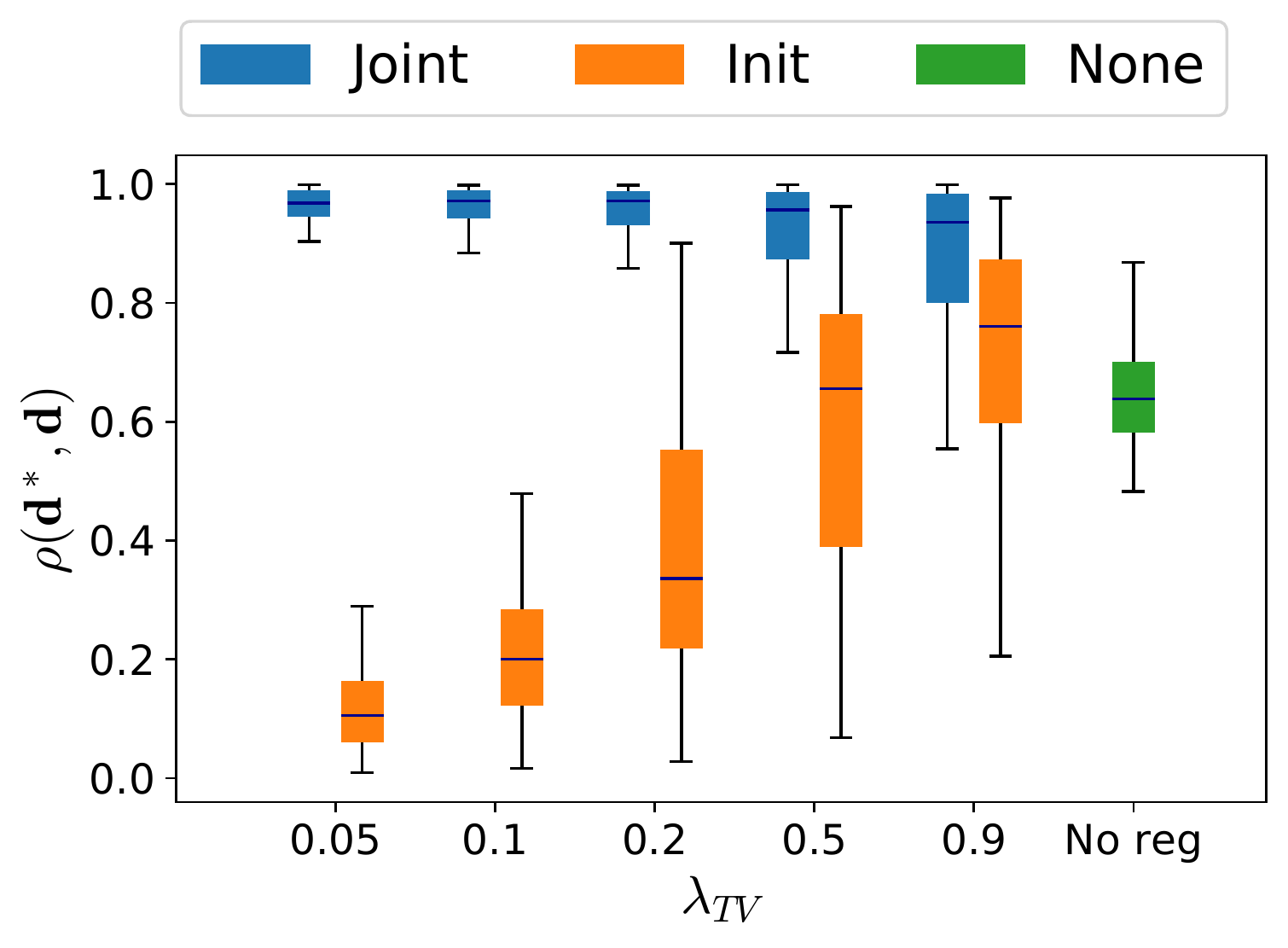}
    \caption{Pearson correlation coefficient between $D$ and the patterns used to generate the signals for different value of $\lambda_{TV}$ and $\lambda=0.5\lambda_{\max}$. The atoms learned without detrending (\emph{green}) are better than the one learned with the addition of a initial detrending (\emph{orange}) while the ones learned with our proposed methods (\emph{blue}) are closer to the ground truth.}
    \label{fig:quantitative}
\end{figure}

\begin{figure*}[htp!]
\begin{subfigure}{\columnwidth}
\centering
  \includegraphics[width=\columnwidth]{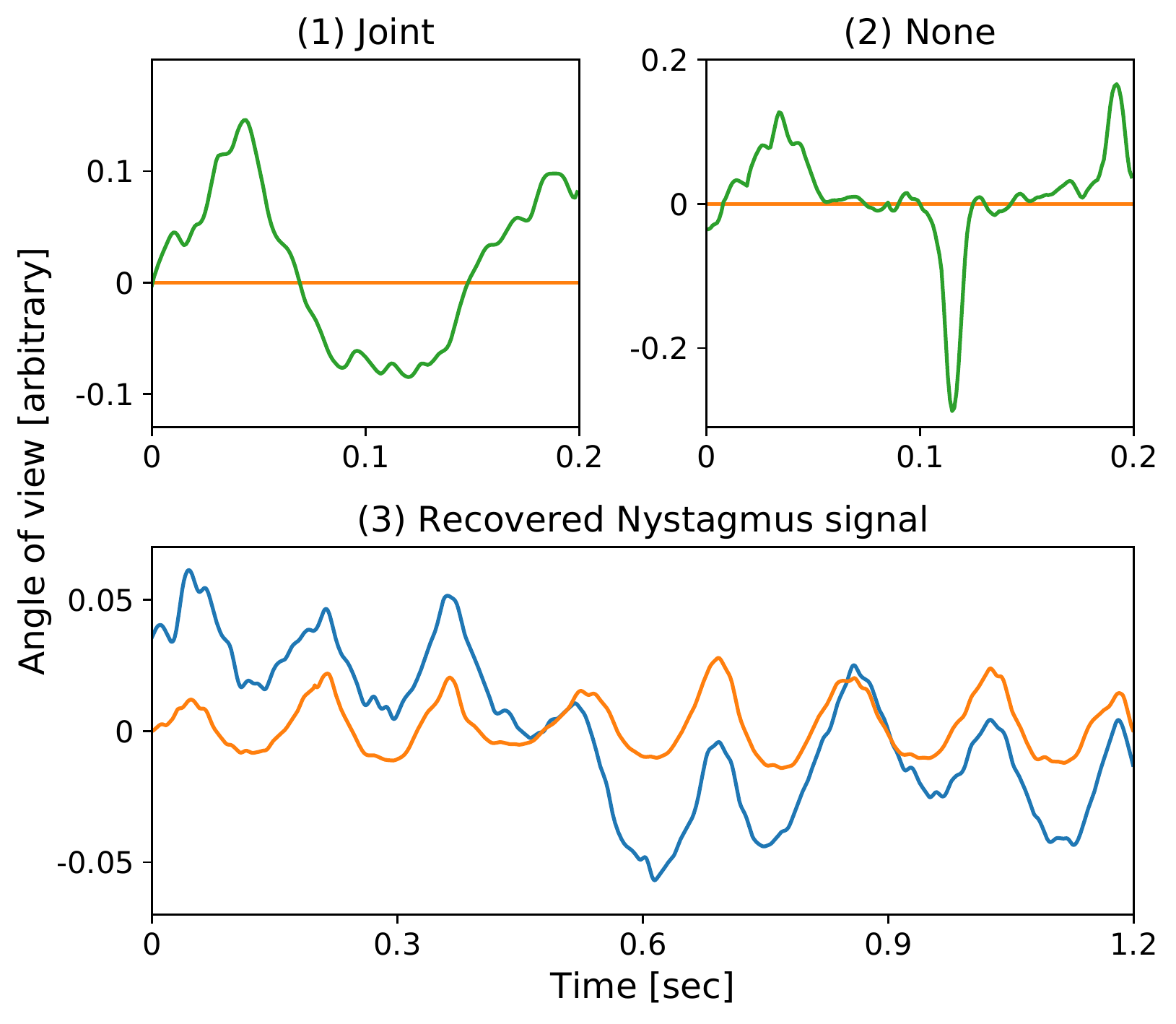}
\caption{Results on Patient 1}
\end{subfigure}
\begin{subfigure}{\columnwidth}
\centering
  \includegraphics[width=\columnwidth]{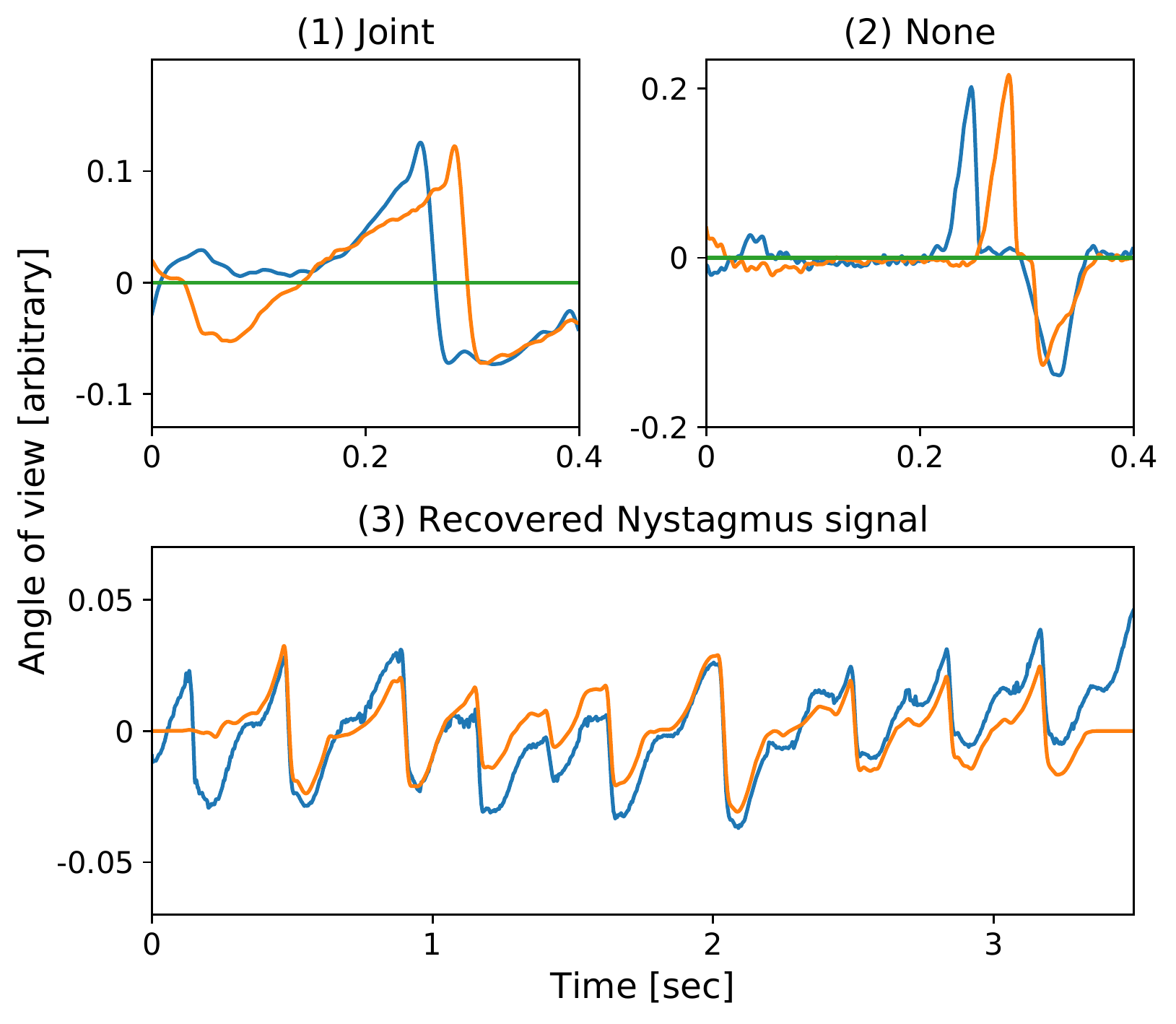}
\caption{Results on Patient 2}
\end{subfigure}
\caption{Results for recordings on two patients presenting (\emph{a}) a SN syndrome with a pendular pattern and (\emph{b})} a INS syndrome with characteristic jerky pattern. The patterns learned with \textsc{Joint} (1) are better at capturing the characteristic waveform of the Nystagmus than the one learned with \textsc{None} (2). The bottom part (3) displays a selected part of the original signals (\emph{blue}) as well as the estimated nystagmic component (\emph{orange}). The signal parts are selected to exclude artefacts and saccades, as it is easier to see the Nystagmus patterns in this configuration.
\label{fig:resultsrealdataset}
\end{figure*}

\autoref{fig:quantitative} summarizes the results for different values of $\lambda_{TV}$ and $\lambda=0.5\lambda_{\max}$. Here, $\lambda_{\max}$ is the smallest value of $\lambda$ for which 0 is solution of \eqref{eq:prob}. \textsc{None} (\emph{green}) is able to recover the generating patterns with a median score of 0.64 in this configuration. When adding the detrending as a preprocessing step, \textsc{Init} (\emph{orange}) is able to better recover the patterns, with a median score of 0.76 if $\lambda_{TV} = 0.9$ is large enough. However, the variance of the recovery is largely increased, indicating that it can either recover well the patterns or fail badly. The proposed method \textsc{Joint} (\emph{blue}) has a median score above 0.95 for $\lambda_{TV} < 0.9$, with a low variance. Note that the behavior of \textsc{Joint} and \textsc{Init} are reverse with the variation of $\lambda_{TV}$. When $\lambda_{TV}$ is small, the initial guess for the trend is very close to the full signal and thus vanilla CSC fails to learn the generating atoms as it is only working on the residual. The \textsc{Joint} method is able to modify the trend estimate to increase the residual where it matters for pattern recovery. As $\lambda_{TV}$ increases, the residual contains more and more of the original signal and the performances of \textsc{Init} are improved toward the ones of \textsc{None}. It becomes better when the trend of a signal is well estimated to not deform the original patterns. But this depends critically on the choice of $\lambda_{TV}$, which cannot be set globally to reach good results.  Also, the benefit of \textsc{Joint} decreases, as the trend is better estimated initially. Thus, the joint estimation of the trend and the patterns allows for an aggressive initial estimate of the trend that will be refined by the joint estimation. We repeated this experiment for different values of $\lambda$ but it did not changed the finding, except by degrading the performances of \textsc{Init}.

\subsection{Results on real data}

We applied \autoref{alg:detrend_csc} to learn 3 patterns from data collected on the two patients presented in \autoref{realdataset}. As suggested by \autoref{fig:quantitative} we tuned the values $\lambda_{TV}=0.05$ and $\lambda=0.9\lambda_{\max}$. The patterns learned with \textsc{Joint} and \textsc{None} are presented in \autoref{fig:resultsrealdataset}. For both patients, the result with vanilla CSC are not highlighting the specific waveform of the Nystagmus while our method is able to capture it.
In the first case, the algorithm extracts a single symmetric pattern, and confirms the pendular nature of the Nystagmus identifyed in the clinical examination.
For the second patient, two similar waveforms are extracted. They are both composed of a slow phase followed by a fast movement, which are closed to text book Nystagmus description \cite{hertle_differential_2013}. The presence of foveation periods in the blue pattern -- stabilization of the movement in the slow phase -- is specific of infantile Nystagmus syndrome. Their presence allows to confirm the INS diagnosis in doubious cases.

\section{Conclusion}
\label{discussion}
Analysing the waveforms of Nystagmus is crucial to the clinician, since the causes  -- and therefore the workup to be performed -- depend on the type of Nystagmus. In many cases, the sole clinical analysis does not allow for a robust classification of the Nystagmus. Sometimes, parents report a transient Nystagmus, which is not visible during the clinic, while infraclinical oscillations are identifyable through oculomotor recording. This can lead to overlooking potentially serious underlying diagnoses, such as retinal dystrophies or brain tumours.

\bibliographystyle{IEEEtran}
\bibliography{root}

\end{document}